\documentclass{article}

\usepackage{amssymb} 
\usepackage{amsmath} 
\usepackage{graphicx} 
\usepackage{epstopdf}

\newtheorem{theorem}{Theorem}
\newtheorem{lemma}{Lemma}

\newtheorem{proof}{Proof}

\begin{document}
%
\title{Modulo-Limiter Modulation of ARMA Processes}

\author{Li Li\footnote{Corresponding author:
li-li@mail.tsinghua.edu.cn}, Yudong Chen, Yi Zhang\\
\\
Department of Automation, Tsinghua University, Beijing, China
100084}

\maketitle

\begin{abstract}
In this paper, we study the quantization errors of modulo
sigma-delta modulated finite, asymptotically-infinite, infinite
causal stable ARMA processes. We prove that the normalized
quantization error can be taken as a uniformly distributed white
noise for all the cases. Moreover, we find that this nice property
is guaranteed by two different mechanisms: the high-enough
quantization resolution \cite{Bennett1948}-\cite{WidrowKollar2008}
and the asymptotic convergence of quantization errors for some
quasi-stationary processes
\cite{ChouGray1991}-\cite{LiChenLiZhang2009}, for different cases.
But the assumption of the smooth density of the sampled random
processes is needed in all the cases.
\end{abstract}

\section{Introduction}
\label{sec:1}

The use of high resolution theory for quantization
error analysis dates back to late 1940s
\cite{ClavierPanterGrieg1947}-\cite{Bennett1948}. In
\cite{Bennett1948}, Bennett demonstrated that under the assumption
of high resolution and smooth density of the sampled stochastic
process, the quantization error can be treated as an additive white
noise. Since then, researcher had further proven the following
conclusion: ``\textit{under most circumstances, the noise is
additively white and uncorrelated with the signal being quantized;
and it is uniformly distributed between minus half a quanta to plus
half a quanta, with a zero mean and a mean square as $\frac{1}{12}$
of the square of a quanta}'' \cite{WidrowKollar2008}. There are
already
some nice surveys in this field, e.g. \cite{Gray1990}-
\cite{WidrowKollar2008}.

In \cite{ChouGray1991}-\cite{ChouGray1992}, Chou and Gray studied
the quantization error that is derived for a modulo sigma-delta
modulator driven by a quasi-stationary stochastic process. They
proved that the quantization errors for a causal
stable MA process behaves just as an additive white noise, if the
sum of the regression coefficients for the MA processes does not
converges to zero.

In a recent report \cite{LiChenLiZhang2009}, we proved that the
conclusion also holds for a fGn process with the Hurst exponent $H
\in (0, \frac{1}{2})$. Such a fGn process can also be viewed as a
special causal stable MA process; while the sum of
its regression coefficients converges to $0$.

Inspired by this new founding, we will further prove in this short
paper that the conclusion holds for any causal
stable MA processes. In the rest of this paper, we will sequentially
discuss the finite, asymptotically infinite and infinite MA
processes, and finally derive the conclusions for the general
causal stable ARMA processes.

\indent

Suppose the original signal $x(n)$ is bounded within $[-b, b]$ in a
finite time horizon $[0, t]$. An $M$-level uniform quantizer in
$[-b, b]$ is applied and the sample rate and the resolution of the
quantizer are high enough.

As shown in \cite{ChouGray1992}, by defining $\Delta =
\frac{2b}{M-1}$, the normalized quantization noise $e(n)$ of $x(n)$
through the modulo-limiter modulator can be written as
\begin{equation}
\label{equ:1} e(n) \triangleq \frac{1}{2} - \left\langle
\sum_{i=0}^{n-1} \left( \frac{x(i)}{\Delta} + \frac{1}{2} \right)
\right\rangle
\end{equation}
\noindent where $\langle x \rangle = x$ mod $1$ is
the fractional part of $x$.

\section{The Results for Finite MA Processes}
\label{sec:2}

For the uniform quantizer, we have the following useful lemma.

\indent

\begin{lemma} \cite{Bennett1948}-\cite{WidrowKollar2008} Suppose
$x(n)$ is a special MA process
\begin{equation}
\label{equ:2} x(n) = z(n)
\end{equation}

\noindent where $z(n)$ is a sequence of 1D random variables having
an identical distribution with a smooth probability density function
(actually $z(n)$ does not need to be independent).
The distribution of the normalized quantization error $e(n)$ under
modulo-limiter modulation converges to the uniform distribution in
$[-\frac{1}{2}, \frac{1}{2}]$ under the assumption of high
resolution. Moreover, the quantization error is additively white and
uncorrelated with the signal being quantized.
\end{lemma}

\indent

The term ``under the assumption of high resolution''
is frequently used in quantization error analysis
\cite{WidrowKollar2008}, \cite{ChouGray1992}. Indeed, it indicates
the existence of a sufficient condition ``under high enough
resolution'', under which the conclusion is true according to the
criteria of uniformity of distribution and whiteness for
quantization errors. Normally, we will further determine which
resolution level is acceptable by numerical testing, with respect to
the practical requirements.

The modulo sigma-delta modulation driven by a input as
Eq.(\ref{equ:2}) can also be viewed as dithering. As pointed out in
\cite{ChouGray1991}, the limit distribution of the normalized
quantization error $e(n)$ after dithering is the uniform
distribution in $[-\frac{1}{2}, \frac{1}{2}]$ regardless of the
distribution of the input signal.

\textit{Lemma 1} immediately leads to the following result for more
general finite MA Processes.

\indent

\begin{theorem} Define a causal stable MA process $x(n)$
\begin{equation}
\label{equ:3} x(n) = \psi(L) z(n) = \sum_{i=0}^{k} \psi_i z(n-i)
\end{equation}
\noindent where $z(n)$ is 1D i.i.d. stochastic
process with a smooth probability density function. $k$ is a
constant, $k \in \mathbb{N}$. If $\psi_i$ does not always equal to
$0$, for $i=1$, ..., $k$, the conclusion in \textit{Lemma 1} also
holds if the other conditions are the same.
\end{theorem}

\begin{proof} Define a process $y(n)$ as
\begin{equation}
\label{equ:4} y(n) = \sum_{i=0}^{k} \psi_i z(n-i)
\end{equation}

Consider the weighted sum of independent random variables
\cite{Petrov1975}-
\cite{Chung2000}, if $\psi_i$ does not always equal to $0$, $y(n)$ is therefore a
sequence of random variables having a certain identical distribution
with a smooth density. Following \textit{Lemma 1}, we can reach the
statement naturally.
\end{proof}

\indent

It should be pointed out that we can allow $\sum_{i=0}^{k} \psi_i =
0$ in \textit{Theorem 1}.

\section{The Results for Asymptotically Infinite MA Processes}
\label{sec:3}

On the other side, we have the following lemma for the
asymptotically infinite MA processes.

\indent

\begin{lemma} \cite{ChouGray1991}-\cite{ChouGray1992} Define a causal stable MA process $x(n)$
\begin{equation}
\label{equ:5} x(n) = \psi(L) z(n) = \sum_{i=0}^{n} \psi_i z(n-i)
\end{equation}
\noindent where $z(n)$ is an 1D i.i.d process having a certain
distribution with a smooth density. if the regression coefficients
$\psi_i$ of this MA process satisfy $\sum_{i=0}^{\infty} \psi_i \ne
0$, then the distribution of the normalized quantization error
$e(n)$ under modulo sigma-delta modulation converges to the uniform
distribution in $[-\frac{1}{2}, \frac{1}{2}]$ under the assumption
of high resolution, when $n \rightarrow \infty$. Moreover, the
quantization error is additively white and uncorrelated
with the signal being quantized.
\end{lemma}

\indent

Based on \textit{Lemma 2}, we will first study a simple cases, where
$z(n)$ has a symmetric stable distribution to illustrate the outline
of our main proof. Then, the results will be extended to the general
MA processes.

\indent

\begin{theorem} Define an causal stable MA process
$x(n)$ satisfying Eq.(\ref{equ:5}), where $\psi_i$ does not always
equal to $0$. If $z(n)$ is an i.i.d stochastic process having a
symmetric stable distribution with the stability index
(characteristic exponent) $\alpha \in [1, 2)$, the
normalized quantization error converges to the
uniform distribution in $[-\frac{1}{2}, \frac{1}{2}]$ under the
assumption of high resolution, when $n \rightarrow \infty$.
\end{theorem}

\begin{proof} According to \cite{Feller1971},
if $z(n)$ is is an i.i.d stochastic process having a symmetric
stable distribution (for simplicity, we assume it is symmetric about
0), the characteristic function of $z(n)$ is written as
\begin{equation}
\label{equ:6} \varphi_z(t) = \textrm{E} \left( e^{i t z} \right) =
\exp \left\{ - \sigma^\alpha \left \vert t \right
\vert^\alpha\right\}
\end{equation}
\noindent where $t$ is the variable of the characteristic function.
$\sigma > 0$ is the scale parameter. When $\alpha \in [1, 2)$, the
distribution function is smooth.

We will discuss three cases in the follows, respectively.

\indent

i) If $\sum_{i=0}^{\infty} \psi_i \ne 0$, according to \textit{Lemma
2}, the conclusion is true.

\indent

ii) If $\sum_{i=0}^{\infty} \psi_i = 0$, but $\psi_i$ does not
always equal to $0$ and $\sum_{i=0}^{\infty} \left \vert \psi_i
\right \vert^\alpha$ converges, we will show that $x(n)$ will
converge to a sequence of random variables having an identical
certain identical distribution with a smooth density, when $n
\rightarrow \infty$.

According to the definition (\ref{equ:5}), we have the limit
characteristic function of $x(n)$ as
\begin{equation}
\label{equ:7} \varphi_x(t) = \textrm{E} \left( e^{i t x} \right) =
\lim_{n \rightarrow \infty} \prod_{i=0}^{n} \varphi_z(\psi_i t)
\end{equation}

Let $\sum_{i=0}^{\infty} \left \vert \psi_i \right \vert^\alpha = S
> 0$, we have
\begin{equation}
\label{equ:8} \varphi_x(t) = \exp \left\{ - \sigma^\alpha \left(
\sum_{i=0}^{\infty} \left \vert \psi_i \right \vert^\alpha \right)
\left \vert t \right \vert^\alpha \right\} = \exp \left\{ -
\sigma^\alpha S \left \vert t \right \vert^\alpha \right\}
\end{equation}
\noindent which indicates that $x(n)$ converges to an identical
symmetric stable distribution \cite{Chung2000}-
\cite{Dudley2002} (Actually, this is \textit{Theorem 9.8.4} shown on
page 328 of \cite{Dudley2002} and a corollary of \textit{Lemma 3}
below). Thus, following \textit{Lemma 1}, $z(n)$ will converge to
the uniform distribution in $[-\frac{1}{2}, \frac{1}{2}]$, when $n
\rightarrow \infty$.

\indent

iii) then, let's consider the situation $\sum_{i=0}^{\infty} \psi_i
= 0$, but $\psi_i$ does not always equal to $0$ and
$\sum_{i=0}^{\infty} \left \vert \psi_i \right \vert^\alpha$ does
not converge.

We will first prove that $\sum_{i = 0}^{\infty} \left \vert \sum_{j
= 0}^{i} \psi_j \right \vert^\alpha$ cannot converge in such a
situation.

For $\alpha \in [1, 2)$, based on the global convexity of $f(x) =
\left \vert x \right \vert^\alpha$ for $x \in \mathbb{R}$, $\alpha
\in [1, 2)$ and Jensen's inequality, we have
\begin{eqnarray}
\label{equ:9} & & 2 \sum_{i = 0}^{\infty} \left \vert \sum_{j =
0}^{i} \psi_j \right \vert^\alpha \nonumber \\
& = & \left \vert \psi_0 \right \vert^\alpha + \sum_{i=1}^\infty
\left( \left \vert \sum_{j=0}^{i} \psi_j
\right \vert^\alpha + \left \vert - \sum_{j=0}^{i-1} \psi_j \right \vert^\alpha \right) \nonumber \\
& \ge & \left \vert \psi_0 \right \vert^\alpha + 2 \sum_{i=1}^\infty
\left \vert \frac{1}{2}\psi_i \right \vert^\alpha \nonumber \\
& = & \frac{1}{2^{\alpha-1}} \sum_{i=0}^\infty \left \vert \psi_i
\right \vert^\alpha + \left[ 1 - \frac{1}{2^{\alpha-1}} \right]
\left \vert \psi_0 \right \vert^\alpha
\end{eqnarray}
\noindent which indicates that $\sum_{i = 0}^{\infty} \left \vert
\sum_{j = 0}^{i} \psi_j \right \vert^\alpha$ also diverges in such
situations.

The limit distribution of the quantization noise can be derived
through the limit of the corresponding characteristic functions. As
shown in \cite{ChouGray1992}, we can rewrite Eq.(\ref{equ:1}) as
\begin{equation}
\label{equ:10} e(n) = 1 - \frac{1}{2} \langle \delta(n) \rangle
\end{equation}
\noindent where $\delta(n) \triangleq \sum_{i=0}^{n-1} \left(
\frac{x(i)}{\Delta} + \frac{1}{2} \right)$.

Notice that $\sum_{i=0}^{\infty} \left \vert \psi_i \right
\vert^\alpha$ does not converge, we can reach the conclusion, if we
then prove that \cite{ChouGray1992}
\begin{eqnarray}
\label{equ:11} \lim_{n \rightarrow \infty} \varphi_{\langle
\delta(n) \rangle} (t) = \left\{
\begin{array}{ll}
1 &, t = 0 \\
0 &, t \ne 0
\end{array} \right.
\end{eqnarray}

The limit characteristic function of $\langle \delta(n+1) \rangle$
can be written as
\begin{eqnarray}
\label{equ:12}
& & \lim_{n \rightarrow \infty} \varphi_{\langle \delta(n+1) \rangle} (2 \pi t) \nonumber \\
& = & \lim_{n \rightarrow \infty} \left \vert \textrm{E} \left\{
\exp
\left[ 2 i \pi t \sum_{i=0}^n \left( \frac{x(i)}{\Delta} + \frac{1}{2} \right) \right] \right\} \right \vert \nonumber \\
& = & \lim_{n \rightarrow \infty} \left \vert \textrm{E} \left\{
\exp
\left[ 2 i \pi \frac{t}{\Delta} \sum_{i=0}^n \sum_{j=0}^i \psi_j z(i-j) \right] \right\} \right \vert \nonumber \\
\end{eqnarray}

The innermost sum in Eq.(\ref{equ:12}) can be grouped as
\begin{equation}
\label{equ:13} \sum_{i=0}^n \sum_{j=0}^i \psi_j z(i-j) = \psi_0 z(n)
+ ... + \left( \sum_{j=0}^{n} \psi_j \right) z(0)
\end{equation}

Therefore
\begin{equation}
\label{equ:14} \lim_{n \rightarrow \infty} \varphi_{\langle
\delta(n+1) \rangle} (2 \pi t) = \lim_{n \rightarrow \infty}
\prod_{i=0}^{n} \varphi_z \left( 2 \pi \frac{t}{\Delta}
\sum_{j=0}^{i} \psi_j \right)
\end{equation}
\noindent where by definition of symmetric stable process, we have
\begin{eqnarray}
\label{equ:15} & & \prod_{i=0}^{n} \varphi_z \left( 2 \pi
\frac{t}{\Delta} \sum_{j=0}^{i} \psi_j \right) \nonumber \\
& = & \prod_{i=0}^{n} \exp \left[ - \sigma^\alpha \left \vert 2 \pi
\frac{t}{\Delta} \sum_{j=0}^i \psi_j
 \right \vert^\alpha \right] \nonumber \\
& = & \prod_{i=0}^{n} \exp \left[ - \frac{\sigma^\alpha \left \vert
2 \pi t\right \vert^\alpha}{\Delta^\alpha} \left \vert \sum_{j=0}^i
\psi_j \right \vert^\alpha \right]
\end{eqnarray}

Noticing that $\sum_{i = 0}^{\infty} \left \vert \sum_{j = 0}^{i}
\psi_j \right \vert^\alpha$ does not converge, given any a small
positive number $\epsilon \in (0, 1)$ and a certain $t \in
\mathbb{R} - \left\{0 \right\}$, we can we can always find a large
enough integer $n^*$ such that
\begin{equation}
\label{equ:16} \sum_{i=0}^n \left \vert \sum_{j=0}^i \psi_j \right
\vert^\alpha \ge - \ln (\epsilon) \frac{\Delta^\alpha}{\sigma^\alpha
\left \vert 2 \pi t \right \vert^\alpha}
\end{equation}
\noindent for $n > n^*$, as $n$ goes to infinity.

Thus, for $t \ne 0$, we have
\begin{equation}
\label{equ:17} \prod_{i=0}^{n} \varphi_z \left( 2 \pi
\frac{t}{\Delta} \sum_{j=0}^{i} \psi_j \right) \le \exp \left( \ln
\epsilon \right) = \epsilon
\end{equation}
\noindent for $n > n^*$, which clearly indicates $\lim_{n
\rightarrow \infty} \varphi_{\langle \delta(n) \rangle} (t) = 0$ for
$t \ne 0$, if we consider (\ref{equ:14}).

On the other hand, we can easily have $\lim_{n \rightarrow \infty}
\varphi_{\langle \delta(n) \rangle} (0) = 1$ by definition.

Therefore, the limit distribution of $e(n)$ is the uniform
distribution among $[-\frac{1}{2}, \frac{1}{2}]$ due to the
definition Eq.(\ref{equ:10}).

The proof for the additively whiteness and non-correlated property
of the quantization error is similar to what had been given in
\cite{Bennett1948}-\cite{LiChenLiZhang2009} and is thus omitted
here. So are the rest.
\end{proof}

\indent

The proof for \textit{Theorem 2} can be extended to the general
cases by using the following useful lemma.

\indent

\begin{lemma} (\textit{L\'{e}vy Continuity Theorem})
\cite{Dudley2002} If $P_n$ are probability laws on $\mathbb{R}^k$
whose characteristic functions $f_n(t)$ converge for all $t$ to some
$f (t)$, where $f$ is continuous at $0$ along each coordinate axis,
then $P_n \xrightarrow[L]{} P$ for a probability law $P$ with
characteristic function $f$.
\end{lemma}

\indent

Based on \textit{Lemma 3}, we have the following general conjecture.

\indent

\begin{theorem} Define a causal stable MA process
$x(n)$ satisfying Eq.(\ref{equ:5}), where $\psi_i$ does not always
equal to $0$. If $z(n)$ is an i.i.d stochastic process having a
certain smooth density function, the normalized
quantization error converges to the uniform distribution in
$[-\frac{1}{2}, \frac{1}{2}]$ under the assumption of high
resolution. Moreover, the quantization error is additively white and
uncorrelated with the signal being quantized.
\end{theorem}

\begin{proof}
If $\sum_{i=0}^{\infty} \psi_i \ne 0$, according to \textit{Lemma
2}, the conclusion is true. In the follows, we will focus on the
cases with $\sum_{i=0}^{\infty} \psi_i = 0$.

Noticing that $\varphi_z$ is bounded as $\left | \varphi_z \right |
\le 1$ and $\prod_{i=0}^{n} \varphi_z(\psi_i t)$ is a monotonic
series in terms of $n$ for any given $t \in \mathbb{R}$, we can see
that $\prod_{i=0}^{\infty} \varphi_z(\psi_i t)$ must converge
pointwise as
\begin{equation}
\label{equ:18} \varphi_x(t) = \lim_{n \rightarrow \infty}
\varphi_{x_n}(t) = \prod_{i=0}^{\infty} \varphi_z(\psi_i t) =
\hat{\varphi}(t)
\end{equation}

Notice that function $\varphi_z$ is sufficiently smooth, for any $t
\ne 0$, we can have the Taylor's expansion of $\varphi_z (\psi_i t)$
around $0$ in Lagrange form as
\begin{eqnarray}
\label{equ:19} \varphi_z( \psi_i t) & = & \varphi_z( 0 ) +
\varphi_z^{'}
(0) ( \psi_i t) + \varphi_z^{''} (\xi_i) ( \psi_i t)^2 \nonumber \\
& = & 1 + \varphi_z^{'} (0) ( \psi_i t) + \varphi_z^{''} (\xi_i) (
\psi_i t)^2
\end{eqnarray}
\noindent where $\xi_i \in [0, \psi_i t]$ if $\psi_i t \ge 0$, or
$\xi_i \in [\psi_i t, 0]$ if $\psi_i t < 0$. Since $\varphi_z^{'}
(0) \sum_{i=0}^\infty ( \psi_i t) = 0$, we have $\psi_n \rightarrow
0$ as $n \rightarrow \infty$, and thus $\xi_n \rightarrow 0$.

Clearly, we have $-1 \le \varphi_z^{'} (0) ( \psi_i t) +
\varphi_z^{''} (\xi_i) ( \psi_i t)^2 \le 0$ due to $\left |
\varphi_z \right | \le 1$.

For any $t \ne 0$, we have
\begin{eqnarray}
\label{equ:20} & & \ln \prod_{i=0}^n \varphi_z( \psi_i t) \nonumber \\
& = & \sum_{i=0}^n \ln \left[ 1 + \varphi_z^{'} (0) ( \psi_i t) + \varphi_z^{''} (\xi_i) ( \psi_i t)^2 \right] \nonumber \\
& = & \sum_{i=0}^n \frac{\varphi_z^{'} (0) ( \psi_i t) +
\varphi_z^{''} (\xi_i) ( \psi_i t)^2}{1 + \eta_i}
\end{eqnarray}
\noindent where we use Taylor's expansion of $f(x) = \ln (1 + x)$ in
the last row.

Here, $\eta_i$ is between $0$ and $\left( \varphi_z^{'} (0) ( \psi_i
t) + \varphi_z^{''} (\xi_i) ( \psi_i t)^2 \right)$. Similarly, we
have $\eta_i \rightarrow 0$, $\varphi_z^{''} (\xi_i) \rightarrow
\varphi_z^{''} (0)$ as $n \rightarrow \infty$.

Given any a small enough non-positive number $x$, we always have $2
x \le \frac{x}{1 + \epsilon x} \le x$, when $\epsilon \in (0, 1)$.
Thus, given a $t \ne 0$, there must exit a large enough $N^* \in
\mathbb{N}$, constants $C_1$, $C_2 \in \mathbb{R}$, $C_3$, $C_4 \in
\mathbb{R}^+$ that
\begin{eqnarray}
\label{equ:21} & & \sum_{i=0}^n \frac{\varphi_z^{'}
(0) ( \psi_i t) + \varphi_z^{''} (\xi_i) ( \psi_i t)^2}{1 + \eta_i} \nonumber \\
& = & \left( \sum_{i=0}^{N^*} + \sum_{i=N^*+1}^n \right)
\frac{\varphi_z^{'} (0) ( \psi_i t) + \varphi_z^{''} (\xi_i) ( \psi_i t)^2}{1 + \eta_i} \nonumber \\
& = & C_1 + \sum_{i=N^*+1}^n \frac{\varphi_z^{'}
(0) ( \psi_i t) + \varphi_z^{''} (\xi_i) ( \psi_i t)^2}{1 + \eta_i} \nonumber \\
& \le & C_1 + \sum_{i=N^*+1}^n \left[ \varphi_z^{'} (0) ( \psi_i t) + \varphi_z^{''} (\xi_i) ( \psi_i t)^2 \right] \nonumber \\
& \le & C_1 + C_2 + \sum_{i=N^*+1}^n \varphi_z^{''} (\xi_i) ( \psi_i t)^2 \nonumber \\
& \le & C_1 + C_2 + C_3 \sum_{i=N^*+1}^n ( \psi_i t)^2
\end{eqnarray}
and
\begin{eqnarray}
\label{equ:22} & & \sum_{i=0}^n \frac{\varphi_z^{'}
(0) ( \psi_i t) + \varphi_z^{''} (\xi_i) ( \psi_i t)^2}{1 + \eta_i} \nonumber \\
& = & C_1 + \sum_{i=N^*+1}^n \frac{\varphi_z^{'}
(0) ( \psi_i t) + \varphi_z^{''} (\xi_i) ( \psi_i t)^2}{1 + \eta_i} \nonumber \\
& \ge & C_1 + 2 \sum_{i=N^*+1}^n \left[ \varphi_z^{'} (0) ( \psi_i t) + \varphi_z^{''} (\xi_i) ( \psi_i t)^2 \right] \nonumber \\
& \ge & C_1 + 2 C_2 + C_4 \sum_{i=N^*+1}^n ( \psi_i t)^2
\end{eqnarray}

Therefore, we have two situations:

\indent

i) if $\sum_{i=0}^\infty (\psi_i)^2$ converges, $\sum_{i=0}^\infty
(\psi_i t)^2$ converges for any a given $t \ne 0$. Based on
Ineq.(\ref{equ:21}), we can see that $\lim_{n \rightarrow \infty}
\ln \prod_{i=0}^n \varphi_z( \psi_i t)$ also converges.

Moreover, given a $t \ne 0$, $\frac{\varphi_z^{'} (0) \psi_i}{1 +
\eta_i}$, $\frac{\varphi_z^{''} (\xi_i) ( \psi_i)^2}{1 + \eta_i}$
are bounded. From Ineq.(\ref{equ:22}), there exist two constants
$C_5$ to $C_8 \in \mathbb{R}$ such that
\begin{eqnarray}
\label{equ:23} & & \ln \prod_{i=0}^\infty \varphi_z( \psi_i t) \nonumber \\
& = & \left( \sum_{i=0}^{N^*} + \sum_{i=N^*+1}^\infty \right)
\frac{\varphi_z^{'} (0) ( \psi_i t) + \varphi_z^{''} (\xi_i) ( \psi_i t)^2}{1 + \eta_i} \nonumber \\
& \ge & C_5 t + C_6 t^2 + \sum_{i=N^*+1}^\infty \frac{\varphi_z^{'}
(0) ( \psi_i t) + \varphi_z^{''} (\xi_i) ( \psi_i t)^2}{1 + \eta_i} \nonumber \\
& \ge & C_5 t + C_6 t^2 + 2 \sum_{i=N^*+1}^\infty \left[
\varphi_z^{'} (0) ( \psi_i t) + \varphi_z^{''} (\xi_i) ( \psi_i t)^2
\right] \nonumber \\
& = & C_7 t + C_8 t^2
\end{eqnarray}

Similarly, we have $\ln \prod_{i=0}^\infty \varphi_z( \psi_i t) \le
C_9 t + C_{10} t^2$ based on Ineq.(\ref{equ:21}), where $C_9$,
$C_{10} \in \mathbb{R}$.

Thus, $\ln \prod_{i=0}^n \varphi_z( \psi_i t) \rightarrow 0$ as $t
\rightarrow 0$. This shows that $\prod_{i=0}^n \varphi_z( \psi_i t)
\rightarrow 1$ and equivalently $\hat{\varphi}(t)$ is continuous
around $0$.

According to \textit{Lemma 3}, if $\hat{\varphi}(t)$ is continuous
around $0$, it is a characteristic function to a certain probability
law. Thus, $x(n)$ converges to a sequence of random variables having
a certain identical distribution with a smooth density. Following
\textit{Lemma 1}, $z(n)$ will converge to the uniform distribution
in $[-\frac{1}{2}, \frac{1}{2}]$, when $n \rightarrow \infty$.

\indent

ii) otherwise, $\sum_{i=0}^\infty (\psi_i )^2$ diverges,
$\sum_{i=0}^\infty (\psi_i t)^2$ diverges for any a given $t \ne 0$,
which indicates $\hat{\varphi}(t)$ is not continuous around $0$.
More precisely, based on Ineq.(\ref{equ:22}), we can easily have
\begin{eqnarray}
\label{equ:24} \lim_{n \rightarrow \infty} \prod_{i=0}^{n}
\varphi_z(\psi_i t) = \left\{
\begin{array}{ll}
1 &, t = 0 \\
0 &, t \ne 0
\end{array} \right.
\end{eqnarray}
\noindent for $t \in \mathbb{R}$.

According to Eq.(\ref{equ:14}), we only need to prove that
\begin{eqnarray}
\label{equ:25} \lim_{n \rightarrow \infty} \prod_{i=0}^{n} \varphi_z
\left( 2 \pi \frac{t}{\Delta} \sum_{j=0}^{i} \psi_j \right) =
\left\{
\begin{array}{ll}
1 &, t = 0 \\
0 &, t \ne 0
\end{array} \right.
\end{eqnarray}
or equivalently
\begin{eqnarray}
\label{equ:26} \lim_{n \rightarrow \infty} \prod_{i=0}^{n} \varphi_z
\left( t \sum_{j=0}^{i} \psi_j \right) = \left\{
\begin{array}{ll}
1 &, t = 0 \\
0 &, t \ne 0
\end{array} \right.
\end{eqnarray}
\noindent to reach the major conclusion.

Based on the global convexity of $f(x) = x^2$ and Jensen's
inequality, we know that
\begin{eqnarray}
\label{equ:27} & & 2 \sum_{i=0}^\infty \left ( \sum_{j=0}^i \psi_j
\right)^2 \nonumber \\
& = & \left \vert \psi_0 \right \vert^2 + \sum_{i=1}^\infty \left(
\left \vert \sum_{j=0}^{i} \psi_j \right \vert^2 +
\left \vert - \sum_{j=0}^{i-1} \psi_j \right \vert^2 \right) \nonumber \\
& \ge & \left \vert \psi_0 \right \vert^2 + 2 \sum_{i=1}^\infty
\left \vert \frac{1}{2}\psi_i \right \vert^2 = \frac{1}{2}
\sum_{i=0}^\infty \left \vert \psi_i \right \vert^2 + \frac{1}{2}
\left \vert \psi_0 \right \vert^2
\end{eqnarray}
\noindent which implies that $\sum_{i=0}^\infty \left( \sum_{j=0}^i
\psi_j \right)^2$ will also diverge if $\sum_{i=0}^\infty (\psi_i
)^2$ diverges.

Therefore, using the similar skills in the proof for \textit{Theorem
2}, we can draw the conclusion based on Eq.(\ref{equ:14}) and
Eq.(\ref{equ:24}).
\end{proof}

\indent

\section{The Results for Infinite MA Processes}
\label{sec:4}

It should be pointed out that \textit{Theorem 3} can be extended to
infinite causal stable MA processes. Actually, we have

\indent

\begin{theorem} Define a causal stable MA process $x(n)$
\begin{equation}
\label{equ:28} x(n) = \psi(L) z(n) = \sum_{i=0}^{\infty} \psi_i
z(n-i)
\end{equation}

The quantization error $e(n)$ also converges to the uniform
distribution in $[-\frac{1}{2}, \frac{1}{2}]$ under the assumption
of high resolution.
\end{theorem}

\indent

The proof is almost the same to that given for infinite cases except

i) we always have $\sum_{i=0}^{\infty} \psi_i$ converges for
infinite cases; otherwise, it is not a well defined MA processes;

ii) Eq.(\ref{equ:14}) is changed to
\begin{eqnarray}
\label{equ:29} & & \lim_{n \rightarrow \infty} \varphi_{\langle
\delta(n+1) \rangle} (2 \pi t) \nonumber \\
& = & \lim_{n \rightarrow \infty} \prod_{i=0}^{n} \varphi_z \left( 2
\pi \frac{t}{\Delta} \sum_{j=0}^{i} \psi_j \right) \cdot
\prod_{i=1}^{\infty} \varphi_z \left( 2 \pi \frac{t}{\Delta}
\sum_{j=i}^{n+i} \psi_j
 \right) \nonumber \\
\end{eqnarray}

Noticing that $\left | \varphi_z \right | \le 1$, we can still apply
the above proof, because we can check $\lim_{n \rightarrow \infty}
\prod_{i=0}^{n} \varphi_z \left( 2 \pi \frac{t}{\Delta}
\sum_{j=0}^{i} \psi_j \right)$ instead.

\indent

Indeed, \textit{Theorem 4} is a general case to the conclusion that
we had drawn for fGn processes with Hurst exponent $H \in (0,
\frac{1}{2})$ in \cite{LiChenLiZhang2009}.

\section{Conclusion}
\label{sec:5}

Based on the theory of ARMA processes
\cite{BrockwellDavis1998}-\cite{BoxJenkinsReisel2008}, we can see
that any a causal stable ARMA process can be
formulated into a corresponding causal stable MA process. Thus, the
above conclusions can be extended to causal stable
ARMA processes.

Reviewing the above discussions, we can find that the nice property
of quantization error is guaranteed by two different mechanisms:
when $\varphi_x(t) = \prod_{i=0}^{\infty} \varphi_z(\psi_i t)$
converges to a continuous characteristic function pointwise, the
asymptotically convergence to uniformly distributed white noise can
be guaranteed by the high-enough quantization resolution
\cite{Bennett1948}, \cite{WidrowKollar2008}; otherwise, this
property is guaranteed by the asymptotic convergence of quantization
errors for certain ARMA processes
\cite{ChouGray1991}-\cite{LiChenLiZhang2009}. But in both cases, the
assumption of the smooth density of the sampled random processes is
needed.

It should also be pointed out that in many applications, the cases
$\sum_{i=0}^{\infty} \psi_i = 0$ are non-trivial. Some interesting
yet important examples can be found in \cite{LiChenLiZhang2009}.


\end{document}